\documentclass{aa}
\usepackage{amsmath}
\usepackage{txfonts}
\usepackage{natbib}
\usepackage{graphicx}
\usepackage{url}

\begin{document}
\title{Extreme Value Theory and the Solar Cycle}

\offprints{A. Asensio Ramos}

\author{A. Asensio Ramos}
\institute{Instituto de Astrof\'{\i}sica de Canarias, 38205, La Laguna, Tenerife, Spain}

\date{Received <date> / Accepted <date>}

\abstract
{}
{We investigate the statistical properties of the extreme events of the solar cycle as measured by the sunspot number.}
{The recent advances in the methodology of the theory of extreme values is applied to the maximal extremes
of the time series of sunspots. We focus on the extreme events that exceed a carefully chosen threshold and
a generalized Pareto distribution is fitted to the tail of the empirical cumulative distribution. A maximum
likelihood method is used to estimate the parameters of the generalized Pareto distribution and confidence
levels are also given to the parameters.
Due to the lack of an automatic procedure for selecting the threshold, we analyze the sensitivity of the 
fitted generalized Pareto distribution to the exact value of the threshold.}
{According to the available data, that only spans the previous $\sim$250 years, the cumulative distribution of the 
time series is bounded, yielding an upper limit of 324 for 
the sunspot number. We also estimate that the return value for each solar cycle is $\sim 188$, while the return
value for a century increases to $\sim 228$. Finally, the results also indicate that the most probable return time 
for a large event like the maximum at solar cycle 19 happens once every $\sim 700$ years and that the probability
of finding such a large event with a frequency smaller than $\sim 50$ years is very small. In spite of the
essentially extrapolative character of these results, their statistical significance is very large.}
{}

\maketitle


\section{Introduction}
When analyzing a given physical process, a large amount of observations opens the possibility of 
applying the power of statistical techniques. The well-developed field of statistics has devised a
panoply of methods that allow us to infer properties from the observed phenomena. Sometimes, these
statistical methods are so powerful that one can extract statistically significant information from
noisy or a reduced set of observations.

One of the most striking examples of this is the case of extreme events. In spite of their inherent 
rarity, extreme events sometimes play important roles and turn out to be of fundamental importance.
In certain fields (analysis of precipitation and floods, maximum temperatures, global climate, etc.),
extreme events are those that produce radical and serious changes. For this reason, the extreme value
theory is well developed in these fields and has been applied during the last decades with great
success.

We can find scarce applications of the theory of extreme values in the field of Astrophysics 
\citep[e.g.,][]{bhavsar_barrow85,bhavsar90,bernstein01}. This situation is somewhat surprising because 
usually the most interesting events that astrophysicist study
are the most extreme ones. In this work, we will apply the theory of extreme values and their most recent
advances to the investigation of the solar activity cycle. One of the best-known indicators of solar
activity is the sunspot relative number. This indicator closely follows the 11-year solar activity cycle and
has been continuously tabulated from $\sim$1750. During the last decades, there has been an increasing
interest in the prediction of the upcoming solar cycles and different techniques have been applied 
\citep[e.g.,][]{li01,orfila02,dikpati06,du06}.
The main cause for this interest has to be found, not only on the pure scientific curiosity of knowing in
advance the amplitude of the following solar maximum, but in the influence of a strong (or weak) solar
maximum on the interplanetary medium. A strong solar maximum can induce solar storms that can
damage the enormous amount of satellites around the Earth. Another reason has to be found on the
feasibility of a tripulated mission to Mars. The required long journey has to be carried out 
away from strong solar maxima in order to minimize the exposure to dangerous doses of radiation.
Our work aims at analyzing the statistical properties of these extreme activity events so that their
frequency and amplitude can be estimated. This could serve to gain some insight on the
efficient amount of shielding needed to protect satellites and/or tripulated missions.

For this reason, having the ability to predict when such extreme solar cycle events would happen is
of interest. A few works have been oriented towards the investigation of the statistics of such
extreme events. \cite{siscoe76a,siscoe76b} used the theory of extreme value to analyze the largest sunspot
number per solar cycle. Later on, \cite{willis_tulunay79} extended the previous works by analyzing
the data from sunspots umbrae, complete sunspots (umbrae+penumbrae) and faculae during nine solar cycles.
Apparently, this kind of analysis has not been repeated during the last 25 years, where more than two solar
cycles have occurred. Additionally, the previous works were also based on older approaches to the statistics 
of extreme values.

We extend in this paper the previous analysis of the solar cycle based on the monthly variation of the
sunspot number. The interest of this work resides in the application of the more recent techniques
to infer the statistical properties of the extreme values of the activity cycle. Our purpose is to
give clues that help us forecast the extremes of the solar cycle and their occurrence frequency for
extended periods of time (not only in the future, but also in the past). In more detail, once the
probability distribution function of extreme events (largest number of sunspots) is characterized,
we investigate whether this distribution is bounded or not and which are the typical events that we
can expect for a given amount of time.

\section{Extreme Value Theory}
It is well-known that the Central Limit theorem \cite[i.e.,][]{feller71} states the asymptotic distribution that a sum
of identically distributed random variables with finite variances
will follow when the number of these variables is sufficiently large. What is
less known is that a similar theory exists for the distribution of the maximum values taken
by a random variable. This apparent lack of awareness has to be found in the fact that, although 
the theoretical formalism is known from several decades, it has taken a long time to devise practical
methods to apply the formalism to real data. As well as the Central Limit theorem deals with the 
functional form of the part of the distribution with largest probability (where the majority of 
the events occur) when a sufficiently large number of random variables with finite variance are summed,
the extreme value
theory deals with the tails of such distributions. As a consequence, since by definition very few events occur
in the tails of the distribution, this lack of observables transforms the estimation of the probability tail
a very difficult task. Fortunately, statisticians have developed efficient methods that can make use of the
few events in the tails to estimate the statistical properties of such extreme events.

Two different approaches are fundamentally used for the analysis of extreme events. The essential difference
is in the way extreme events are defined. Let $\{X_i,i=1\ldots n\}$ be a sequence of independent random variables
that have a common distribution function $F$. Each $X_i$ can be considered as a measure of a random process
taken with a certain timestep $\Delta t$. The first approach, termed the \emph{block maxima} approach \citep{fisher_tippett28,coles01}, takes as extreme events the 
maximum (or minimum) value of the random variable in fixed intervals of time. For instance, if the timestep
is considered to be one hour, the maximum among 24 consecutive measurements is the daily maximum. 
The second approach, termed the 
\emph{peaks over threshold} (POT) approach \citep[e.g.,][]{coles01}, takes as extreme events all the values 
of the time series that 
exceed a given threshold. The first approach was the one taken by \cite{siscoe76a,siscoe76b,willis_tulunay79}
for the analysis of extreme values in the temporal evolution of sunspot numbers. It has been selected 
traditionally for the analysis of time series in which a clear seasonal (periodic) behavior is detected.
However, in spite of the theoretical simplicity of the block maxima approach, it suffers from important drawbacks.
One of the most important limitations is that it tends to make an inefficient use of the data. The importance
of the necessity to overcome this lack of efficiency lies in the fact that the extreme value theory deals with
extreme events, which are, by definition, scarce. For this reason, POT is becoming the 
method of choice in recent applications of the theory of extreme events, fundamentally for the efficient
use of the reduced amount of data available. 

\begin{figure}
\resizebox{\hsize}{!}{\includegraphics{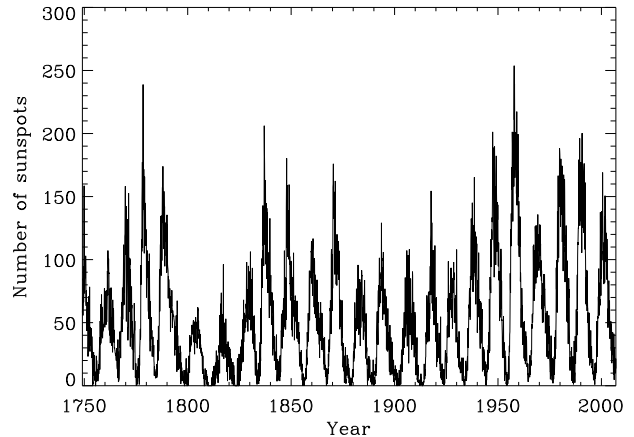}}
\caption{Solar cycle variation of the International Sunspot Number. In order to decrease the dispersion,
we focus on the monthly averaged value.}
\label{fig:data}
\end{figure}

We briefly present the theoretical results that we use in this paper. Assume that we measure a random variable
at constant time intervals and that we obtain a sequence $\{X_i,i=1\ldots n\}$. The measurements have to
be independent and they have a common distribution function. Then, the sequence can be
described with the aid of the cumulative distribution function $F$. Since we are focusing on
extreme events, we will be interested in the tail of such distribution. The POT approach is based on
analyzing what is known as the \emph{conditional excess distribution function}, $F_u(y)$, defined as:
\begin{equation}
F_u(y) = P(y \geq X-u \, | \, X > u), \qquad 0 \geq y < \infty,
\end{equation}
where $X$ is the random variable, $u$ is the threshold used to distinguish the maximum values and we have 
assumed (for simplicity)
that $X$ can take infinitely large values. This $F_u(y)$ function therefore describes the
cumulative probability that, given a value of the random variable larger than the threshold, it
exceeds the threshold by a quantity $y$. In the case that the complete cumulative distribution function 
$F$ is perfectly known, $F_u(y)$ would also be known. In realistic applications, the cumulative distribution
has to be empirically estimated and the tail of the distribution is often poorly sampled. For this reason, the
estimation of $F_u(y)$ is usually not possible or uncertain. Theoretically, there is a relation between
$F$ and $F_u(y)$:
\begin{equation}
F_u(y) = \frac{F(u+y)-F(u)}{1-F(u)}, \qquad y>0.
\label{eq:conditional_excess}
\end{equation}
The feasibility of the POT approach finds its roots in the powerful theoretical result by 
\cite{pickands75}, who derived that, for a very large class of underlying $F$ 
distributions\footnote{According to \cite{pickands75}, the class of distributions $F$ that fulfill the 
theorem are those whose (block) maxima follow one of the three families of extreme value 
distributions \citep{fisher_tippett28}. These
are the Gumbel, Fr\'echet or Weibull distributions.}, the 
conditional excess distribution function is well approximated by the generalized Pareto distribution (GPD):
\begin{equation}
F_u(y) \approx 
\begin{cases}
1 - \left( 1+ \frac{\xi}{\sigma} y \right)^{-1/\xi}  & \text{if $\xi \neq 0$} \\
1 - e^{-y/\sigma} & \text{if $\xi = 0$},
\end{cases}
\end{equation}
with $y \in [0,\infty]$ if $\xi \geq 0$ and $y \in [0,-\sigma/\xi]$ if $\xi < 0$. 
The GPD is a general
cumulative distribution function that is able to model the behavior of different tails depending on the 
exact value of the parameters. The quantity $\xi$ gives information about the shape, in particular, the
``strength'' of the tail. An
exponential-type distribution (normal, exponential, log-normal) is found for $\xi=0$, a bounded 
beta-type distribution (beta or uniform) is found for $\xi < 0$ (zero probability
is assigned for events above a certain limit) while a heavy-tailed Pareto-type distribution (power law, 
Pareto, Cauchy) is found for $\xi >0$ (the tail falls slower than an exponential). The quantity 
$\sigma$ gives information about the scale of the distribution. Although
not applied in this work, it is important to remind that there is a duality
between these distributions and the extreme value Gumbel, Fr\'echet and Weibull distributions of the block
maxima approach. 

\begin{figure*}
\resizebox{\hsize}{!}{\includegraphics{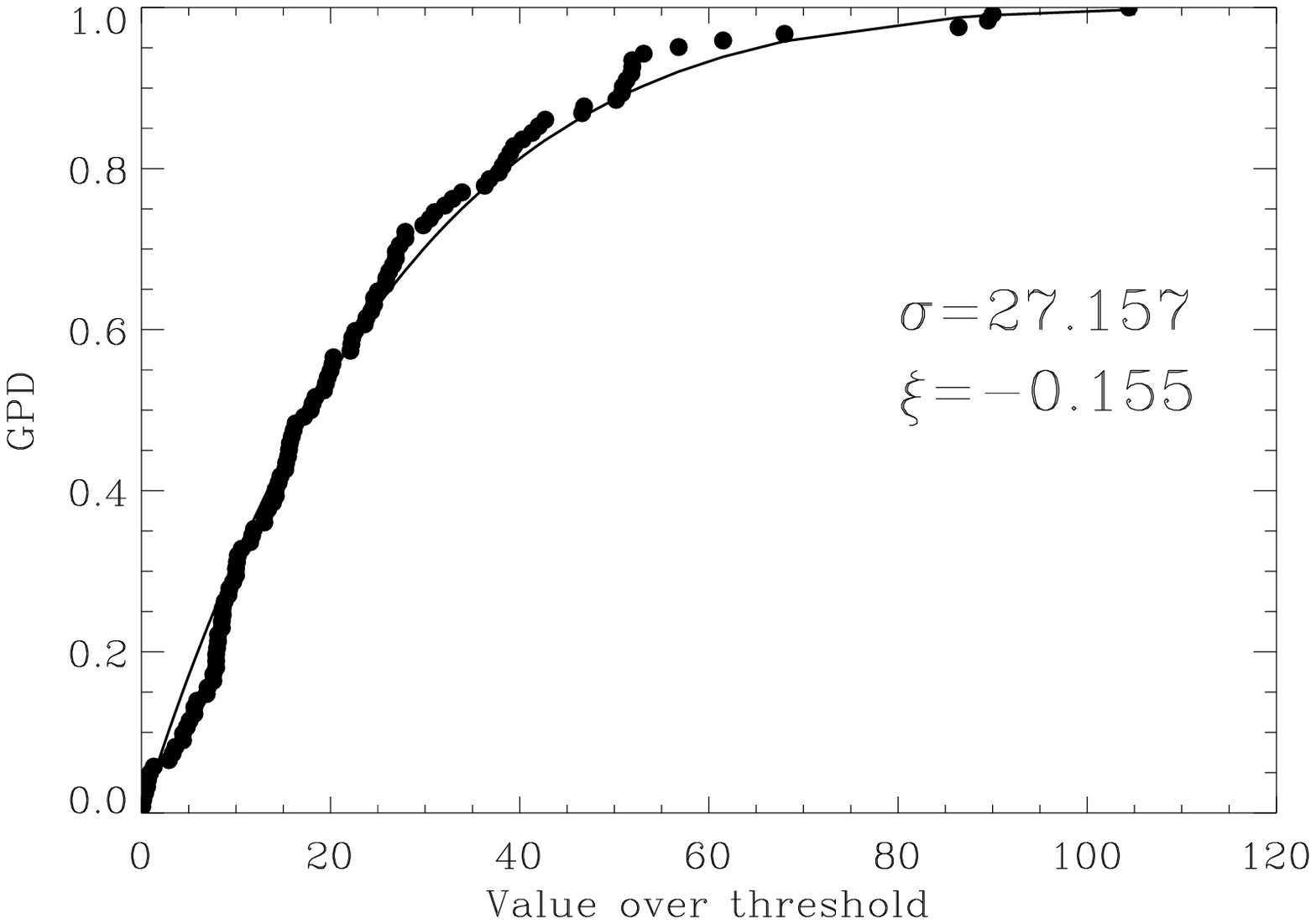}%
\includegraphics{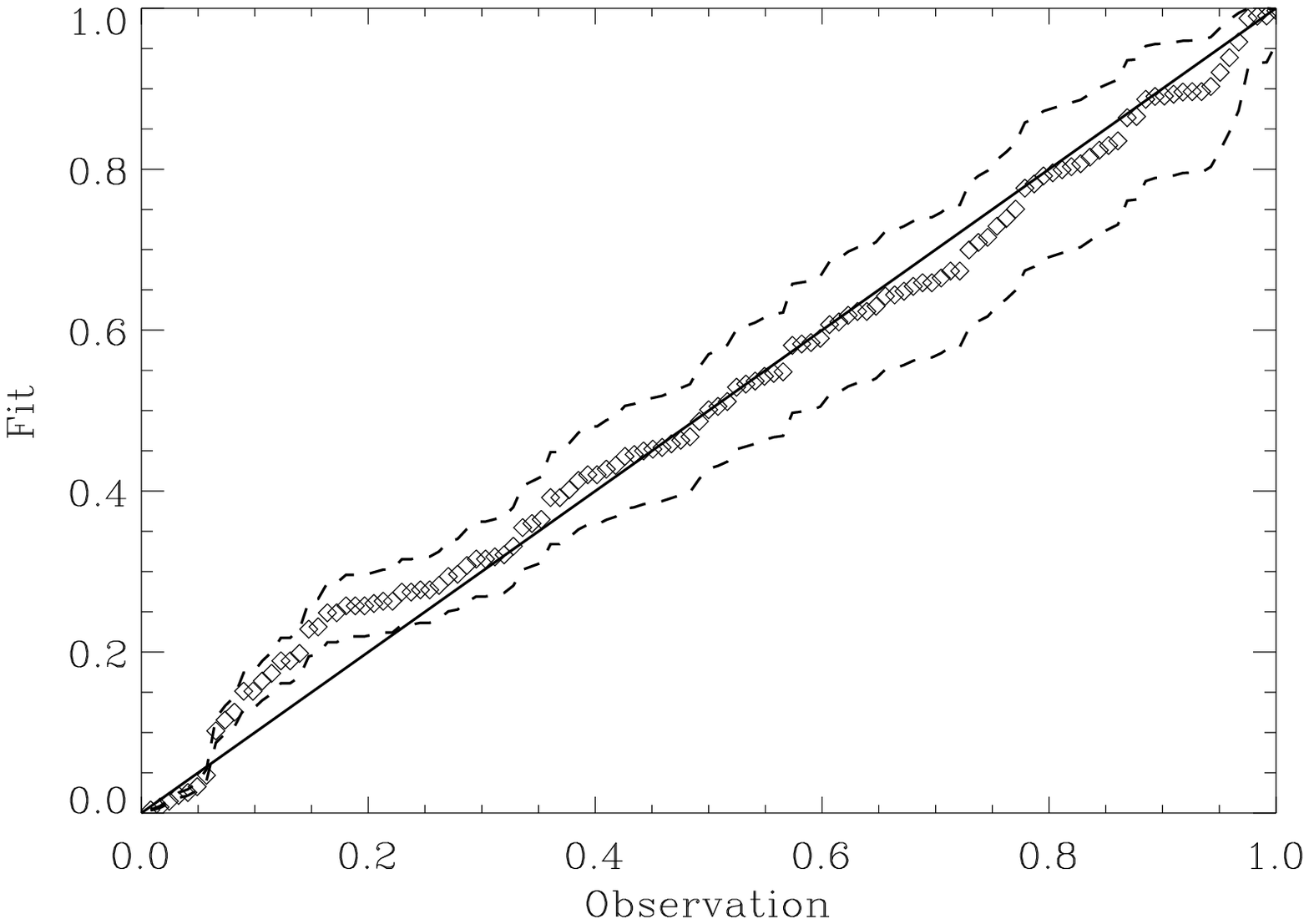}}
\caption{Summary of the quality of the fit of the empirical cumulative distribution to a generalized Pareto
distribution. The left panel shows the value of the cumulative distribution for different values of the
sunspot number above the selected threshold of 149.4 (that leaves only 4\% of the points on the time series
with larger values) together with the GPD fit. The right panel shows the variation of the quality of
the fit when the uncertainties in the $\sigma$ and $\xi$ parameters of the GPD are taken into account. It is
clear that the GPD seems to produce a good representation of the empirical cumulative distribution.}
\label{fig:gpd_fit}
\end{figure*}

With the aid of the theorem developed by \cite{pickands75}, the functional form of the cumulative distribution
function for events above $u$ can be written. Making $x=u+y$ and solving for $F(x)$ in Eq. (\ref{eq:conditional_excess}), 
the cumulative distribution function for events above the threshold $u$ can be written as:
\begin{equation}
F(x) = 1- \frac{N_u}{n} \left[ 1+\frac{\xi}{\sigma} (x-u) \right]^{-1/\xi}.
\label{eq:cumulative_distribution}
\end{equation}
The previous expression assumes that the value of the cumulative distribution at $u$ is given
by the estimation $(n-N_u)/n$ (the so-called historical simulation), with $n$ the total number of 
points in the time series we are analyzing and $N_u$ the number of points above the threshold. This estimation 
is expected to be accurate if the threshold is high enough. Obviously, the functional form
described by Eq. (\ref{eq:cumulative_distribution}) is only valid for $x \geq u$. Once the function
$F$ is known, all the statistical properties of the extreme events can be calculated. One of the
most interesting statistical properties is the so-called ``return time''. This is defined as the
typical time that one has to wait until an event of amplitude $x_\mathrm{ret}$ happens again. It can
be estimated from Eq. (\ref{eq:cumulative_distribution}) by setting $x=x_\mathrm{ret}$ and making
the identification $t_\mathrm{ret}^{-1}=1-F(x_\mathrm{ret})$. The units of this time variable depend on
the units of the timestep $\Delta t$.

The parameters of the GPD are usually obtained from the empirical data by means of a maximum likelihood
estimation. Assuming that $y_1,y_2,\cdots, y_N$ are the $N$ values of the original time series that
exceed the threshold $u$, the log-likelihood is \cite{coles01}:
\begin{equation}
\ell(\sigma,\xi) = -N \log \sigma - \left( 1+\frac{1}{\xi} \right) \sum_{i=1}^{N} \log \left( 1+ \frac{\xi y_i}{\sigma} \right).
\label{eq:log_likelihood}
\end{equation}
Since the value of the $\sigma$ and $\xi$ parameters that maximize the log-likelihood cannot be found analytically,
the $\ell(\sigma,\xi)$ function is maximized using standard numerical optimization methods 
\citep[e.g.,][]{numerical_recipes86}. Once the maximum likelihood value of the parameters are found, it is 
possible to calculate confidence intervals. A standard method relies on the assumption that the likelihood
function is given approximately by a normal distribution \cite{coles01}. Under this assumption, the confidence interval
can be estimated with the aid of the estimated curvature of the likelihood function, which is proportional
to the Hessian matrix of Eq. (\ref{eq:log_likelihood}). A more refined method, and the one that we use in this
paper, is to use the information from the likelihood function itself. This allows us to give asymmetric
confidence intervals in the parameters due to the skewness of the likelihood function.

\section{Application to the Solar Cycle}
The time series representing the solar activity cycle during the last 257 years is shown in Fig. \ref{fig:data}. The
data represents the sunspot number, that is an estimation of the number of individual sunspots (counting individual
sunspots and group of sunspots). The data has been tabulated since 1750 and it is nowadays known as the
International Sunspot Number\footnote{\url{http://sidc.oma.be/}}. The time series presents a clear regularity
with time as a consequence of the influence of the solar cycle on the surface magnetism. Under these
circumstances, the premises on which the POT formalism lies are not fulfilled. In particular the
random variables are not independent because there is a certain degree of correlation between consecutive
events: a large sunspot number is typically followed by another large sunspot number. Several techniques
have been devised to overcome this difficulty. One of the most applied ones is the ``de-clustering'' of the 
time series \cite[e.g.,][]{coles01}. The method consists on locating clusters in the excedance over the 
threshold and representing them by the maximum value inside each cluster. This has two undesired 
consequences: (i) the number of events available for the GPD analysis is reduced and (ii) a somewhat
arbitrary criterion for the cluster definition has to be included. Recently, \cite{fawcett_walshaw06} have
shown that this de-clustering technique introduces biases in the maximum likelihood estimations of
$\sigma$ and $\xi$. They also show that the direct application of the POT method using the whole time
series neglecting any temporal periodicity leads to negligible biases. The price to pay is that the
confidence intervals for the GPD parameters are larger than those obtained using standard techniques
\cite{coles01}. Following \cite{fawcett_walshaw06}, we apply the POT method to the sunspot number time series 
without any de-clustering technique. The POT formalism also requires the underlying distribution
of the random variables to be stationary \citep[see, e.g.,][]{coles01}. The large amount
of works that are successful in reproducing the time evolution of the International Sunspot Number 
using deterministic methods \citep[e.g.,][]{verdes00,dikpati06,choudhuri07,cameron07}
suggest that this is the case. Therefore, we can safely consider that 
the physics (probability distribution function of the random variables) driving the solar cycle does not 
vary with time.

\begin{figure}
\resizebox{\hsize}{!}{\includegraphics{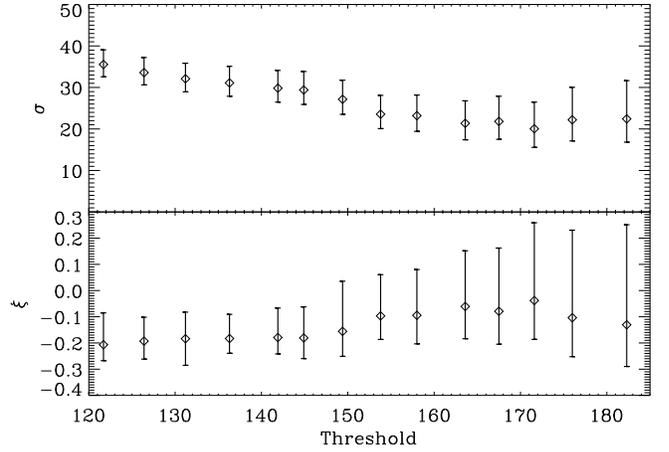}}
\caption{Value of the $\sigma$ and $\xi$ parameters obtained for different values of the 
threshold. When the threshold is increased, the standard deviation of the errors in the
parameters increases due to the decrease in the number of available extreme values. On the contrary,
when the threshold is decreased, the error bars of the parameters decrease. However, a trade-off has
to be found so that the threshold is made large enough so that the GPD is a good approximation of the
empirical distribution but not too large so that the parameter estimation is corrupted by 
poor sampling effects. An intermediate value of 149.4 (4\% of the points lie above this value) seems
to represent a good compromise.}
\label{fig:fits_several}
\end{figure}

\begin{table}
\caption{Parameters and return level estimates. The values are obtained for a threshold of $u=149.4$.}             
\label{tab:value_fit}      
\centering                          
\begin{tabular}{c c c}        
\hline\hline                 
Parameter & Estimate & 95\% confidence interval  \\    
\hline                        
   $\sigma$ &  27.157 & [23.525,31.745]  \\      
   $\xi$    & $-0.155$  & [-0.251,0.035] \\
   $-\sigma/\xi+u$ & 324.0 & [270.0, 990.0] \\
11-year return level & 187.8 & [180.2,203.0] \\
100-year return level & 228.4 & [208.1,253.8] \\
\hline                                   
\end{tabular}
\end{table}

In this paper we only focus on the statistics of the upper tail of the distribution, i.e., maximal values
of the sunspot number. Although the application of the POT formalism to the analysis of the distribution of minimal 
values is also possible, larger time series are needed. We briefly discuss this issue in \S\ref{sec:conclusions}.
In order to apply the POT formalism, a threshold $u$ has to be fixed. The threshold has to be sufficiently large so that 
the generalized Pareto distribution is a suitable functional form for describing the tail of the cumulative distribution
and it has to be sufficiently small so that enough values are available to give an accurate estimation
of the parameters of the GPD. There is not any known automatic procedure for the selection of the threshold.
In this paper we choose a value of the threshold based on reasonable ideas and we verify which is the behavior of
the parameters of the GPD for different values of the threshold. In our case, $u$ has been chosen as the value that
leaves 96\% of the points of the time series below and only 4\% of the points are considered as extreme values. For
the dataset shown in Fig. \ref{fig:data}, we find $u=149.4$. From the original set of 3096 data points, we leave
122 points above the threshold which are used to fit the GPD neglecting any time dependence. The empirical 
cumulative distribution function for points
above the threshold is built and the values of $\sigma$ and $\xi$ that give the best fit are obtained.

The GPD parameters have been estimated maximizing the log-likelihood given by Eq. (\ref{eq:log_likelihood}). As
noted above, such an approach permits to obtain the most probable values and their confidence intervals.
In our case, we obtain $\sigma=27.157^{+4.587}_{-3.632}$ and
$\xi=-0.155^{+0.19}_{-0.09}$, as shown in Table \ref{tab:value_fit}. With these values, the ensuing cumulative distribution
function is shown in the left panel of Fig. \ref{fig:gpd_fit}, where we show the value over the threshold 
on the horizontal axis
and the value of the GPD on the vertical axis. The right panel of Fig. \ref{fig:gpd_fit} shows the 
empirical cumulative distribution versus the fit $F_u(y)$ using the GPD. This is the 
so-called probability plot, and it clearly indicates that the GPD produces a good approximation to the empirical cumulative
distribution. Several interesting points deserve comment. Firstly, the confidence intervals presented in Table
\ref{tab:value_fit} are probably underestimating the true confidence intervals because we have neglected the
temporal dependence. Secondly, a negative value for the shape
parameter $\xi$ is found. The evidence for this is strong because the 95\% confidence interval is 
almost exclusively in the negative domain. According to the results
presented before, this suggests that the cumulative distribution 
is bounded and that zero probability is assigned to excedances above a certain limit, given by the ratio 
$x_\mathrm{lim}-u=-\sigma/\xi$.
In our case, we find that the fitted GPD assigns zero probability to extreme values larger than 175.6 above
the threshold. Taking into
account the chosen threshold, this gives a limit of $x_\mathrm{lim}=324.0$. This value is consistent with the 
data presented in Fig. \ref{fig:data}. Note also that the 95\% confidence interval is $[270.0, 990.0]$, which
has been obtained from the
likelihood function. It gives a very stringent value of the lower limit (as obvious because of the presence of 
a large amount of data below the threshold) but a very large upper limit. A 68\% confidence interval 
is estimated to be $[288,420]$. This is a clear indicative of the asymmetry of the likelihood function with a 
very long tail towards larger $x_\mathrm{lim}$, giving the idea that the information for a strong upper limit is 
hardly present in the data. However, it is important to take into account that these results can
fluctuate depending on the exact value of the chosen threshold. Furthermore, it can also fluctuate in the
future if the same calculation is carried out with more data from subsequent solar cycles. In order to
analyze the strength of this conclusion, we have carried out the fit of the cumulative distribution for 
different values of the threshold. The results are shown in Fig. \ref{fig:fits_several} for different values of the
threshold $u$, the upper panel showing the values obtained for $\sigma$ and the lower panel the values for $\xi$. 
If the GPD is a reasonable model for the excedances above a certain threshold $u_0$, it should remain 
reasonable for larger values of the threshold \citep[see, e.g.,][]{coles01}. As a consequence, the estimates
of $\sigma$ and $\xi$ should remain constant above $u_0$ provided $u_0$ is a valid threshold. We see that this
happens in our case for $u > 150$, the value we have chosen above.
It is interesting to note that for $u \sim 120$, approximately $9$\% of the points lie above the threshold, while for
$u \sim 180$, less than 1\% of the points lie above. There is a clear increase on the uncertainty of the
retrieved parameters when the threshold increases due to the lack of points. What seems clear is that, 
for $u < 150$, the $\xi$ parameter has only negative values inside the error bars. For larger values of the
threshold, both negative and positive values for $\xi$ are obtained, although negative values tend to be
more probable. Focusing now on the threshold $u=149.4$ and taking into account the uncertainty in the fitted 
parameters, we find that the upper limit of the tail can be found inside the interval $(217,430)$, as indicated in 
Table \ref{tab:value_fit}.

\begin{figure}
\resizebox{\hsize}{!}{\includegraphics{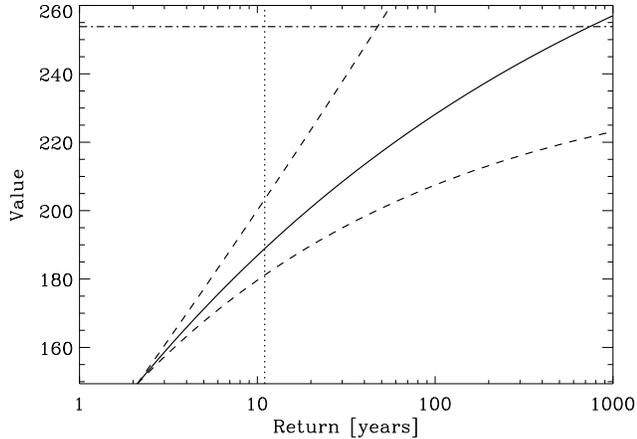}}
\caption{Return value for a given amount of time (or return time for a given level) for the fit obtained with 
a threshold of $u=149.4$. The dashed lines indicate the confidence levels of the return value taking into
account the uncertainty in the GPD parametrization. The vertical dotted line indicates 11 years, approximately
one solar cycle. The horizontal line indicates the value of the maximum found on solar cycle 19.}
\label{fig:time_return}
\end{figure}

Once the cumulative distribution given by Eq. (\ref{eq:cumulative_distribution}) is obtained, several statistical
properties of the extreme values can be inferred. One of the most interesting in terms of prediction of future
events are the so-called return times and return level. The return time is the typical time one has to wait
until an event reaches and surpasses a threshold. Similarly, the return level is the typical extreme event
one would find after waiting for a given amount of time. These quantities are obtained easily from 
Eq. (\ref{eq:cumulative_distribution}) and they are shown in Fig. \ref{fig:time_return}. For consistency, 
we only show the results for values above the chosen threshold. The dashed lines present the confidence interval
that are induced by the uncertainty in the inferred parameters of the GPD. The vertical dotted line 
approximately indicates a solar cycle, of the order of 11 years. For such amount of time, the return level
equals to $\sim 187$. If we take into account the confidence interval, we find that the return level lies
in the interval $[180,203]$. These values appear to be consistent with the empirical results from the past, according 
to Fig. \ref{fig:data}. This is produced by the similarity between the empirical tail of the extreme distribution
and the GPD, that makes the statistical properties inferred from the GPD a good estimation of the statistical
properties of the empirical distribution. If we assume that the previous 23 solar cycles are 
representative of the behavior of the
Sun during a longer time and unless a long-period modification of the solar cycle exists, the previous estimations
of the return levels are statistically significant. For the case of the return level for 100 years, we find
a value of $\sim 228$, with a confidence interval of $[208,254]$. Again, this value is consistent with the
data. 

Concerning the return times, it is interesting to estimate them for the most extreme cases in the observed
dataset. It is important to have in mind that relying on the GPD for very long-term extrapolation can
be extremely dangerous. The dataset on which we have applied the extreme value theory spans only $\sim 250$ years,
so that one should not blindly rely on the return values for large events if they only happen once in
the original data. For instance, according to the GPD, an event like the extremely strong peak of cycle 19 
(the maximum of 253.8 took place during 1957) occurs 
once every $\sim 700$ years. Taking into account the uncertainty in the GPD fit, we find that it is possible to
give only a lower limit to this return time because the upper limit is unbounded. The horizontal line
indicates the maximum value of 253.8 obtained during cycle 19. The intersection of this horizontal line
with the solid line given by the GPD extrapolation occurs at $\sim 700$ years, showing the
most probable value of the return time. The
intersection with the dashed lines indicate the intervals of 95\% confidence. In this case, we find that 
the GPD extrapolation implies that such an extreme event happens, with 95\% probability, with a
frequency above once every $\sim 50$ years, with an apparently unbounded upper limit.

We want to leave a word of caution on the values obtained above because of their extrapolative character. 
However, we also want to stress the fact that this extrapolative character is based on strong theoretical 
roots. A small variation on the results obtained in this work might be expected as soon as more data showing more 
extreme events is available. This variation with respect to the values presented above should be very small
if the underlying probability distribution function (essentially, the physics that drives the solar cycle)
that we have calculated does not change appreciably with time.

\section{Concluding remarks}
\label{sec:conclusions}
We have presented an extreme value theory analysis of the solar cycle as measured by the sunspot number. The 
peak-over-threshold method has been applied. The analyzed time series presents a certain degree of correlation
because a large
number of sunspots is usually followed by another large number of sunspots and these variables cannot be considered
to be uncorrelated. Following \cite{fawcett_walshaw06}, we have applied the POT method to the time series without
any de-clustering technique. As a consequence, the confidence intervals presented in this work could be an 
underestimation of the
real confidence intervals because the likelihood function is affected by the presence of time correlation in the
time series. Our results indicate that the distribution of extreme solar cycle events is bounded so that the
value of 324 cannot be exceeded. The analysis of the confidence intervals give that, with 95\% confidence,
this maximum value is larger than 288. Of more relevance are the 11-years and 100-years return values. The results
indicate that there is a very high probability of finding values in the range $[180,203]$ every solar 
cycle, and values in the range $[208,254]$ every $\sim$10 solar cycles. Additionally, we have shown that an extreme
event like that on solar cycle 19 (during 1957) occurs with a frequency above once every $\sim$50 years.

The results obtained in this paper are based on the statistical analysis of the tail of 
the sunspot number distribution. This analysis is driven by the extreme value theory that is based on
strong theoretical roots developed during the last 50 years. Such application relies on the assumption
that the limiting behavior of the stochastic process behind the observed time series can be obtained from
the application of certain mathematical limits. Then, a functional form for the tails of distributions is available
and we only have to fit this tail distribution to our dataset. However, this approach has limitations. One 
of the strongest is that it is not yet clear whether the limiting mathematical models that we have used 
can be directly applied to finite time series. It is expected that, in the limit of an infinitely large
time series, the models correctly reproduce the tails of the underlying distribution. However, when the time 
series is of limited size, fluctuations can be of importance and lead to inaccuracies. In our case, 
the results that we present appear to support the fact that the statistics of extreme events are correctly 
reproduced under the framework of the extreme value theory. The fact that the theory explains the already
observed extreme events is also favorable. The validity of the theory is also supported
by the large amount of practical applications of the theory that we found in the literature.

The extreme value theory can also be applied to minima. One of the most interesting future applications of this
approach is to estimate the return time for long low-activity periods, with the aim of estimating the 
frequency with which extreme events like the Maunder minimum may take place. A much longer time series of
solar activity would be needed for such an estimation. Furthermore, an appropriate re-definition of the 
time series has to be carried out in order to transform these low-activity periods into extremes cases. 
A possibility that could be of interest is to use the solar sunspot number reconstructed from 
\element[][14]{C} activity during the last 11000 years \citep{solanki04}. This reconstruction shows
periods of low activity similar to the Maunder minimum whose probability (and ensuing
return time) could be calculated in the framework of the extreme value theory.

\begin{acknowledgements}
\thanks{
I thank A. L\'opez Ariste, R. Manso Sainz and V. Mart\'{\i}nez Pillet for helpful discussions
and the anonymous referee for some constructive remarks.
This research has been funded by the Spanish Ministerio de Educaci\'on y Ciencia through project AYA2004-05792. }
\end{acknowledgements}



\end{document}